\documentclass[12pt,a4]{article}

\usepackage{amsmath}
\usepackage{graphicx}
\usepackage[table]{xcolor}
\usepackage{booktabs}
\usepackage{amssymb,bm}
\usepackage{textcomp}
\usepackage{multirow}
\usepackage{subcaption}
\usepackage{adjustbox}
\usepackage{url}

\begin{document}
	
	\title{Text-dependent Speaker Verification (TdSV) Challenge 2024: Challenge Evaluation Plan}
	
	\author{Hossein Zeinali, Kong Aik Lee, Jahangir Alam, Luka\v{s} Burget}
	
	\date{Version 1.1, April 20, 2024}
	
	\maketitle
	
	\section{Introduction}

	This document outlines the Text-dependent Speaker Verification (TdSV) Challenge 2024, which centers on analyzing and exploring novel approaches for text-dependent speaker verification. The primary goal of this challenge is to motive participants to develop single yet competitive systems, conduct thorough analyses, and explore innovative concepts such as multi-task learning, self-supervised learning, few-shot learning, and others, for text-dependent speaker verification.

    Building upon the achievements of the Short-duration Speaker Verification (SdSV) Challenge 2020 and 2021, the TdSV Challenge 2024 focuses exclusively on text-dependent verification in two distinct scenarios. The first scenario involves conventional TdSV, while the second track entails speaker enrollment using user-defined passphrases. The evaluation dataset utilized for the challenge is derived from the second version of the versatile DeepMine dataset \cite{deepmine2018odyssey, deepmine2019asru}. For this challenge, Parts 1 and 3 of the dataset are employed. The subsequent section provides a comprehensive description of both tasks.
		
	\section{Tasks Description}
	
	\subsection{Task 1 - Text-Dependent Speaker Verification}
	
	Task 1 of the TdSV Challenge 2024 involves speaker verification in the conventional text-dependent mode: determining whether a specific phrase was spoken by the target speaker, given a test segment of speech, and the target speaker's enrollment data. Unlike text-independent speaker verification, Task 1 considers the lexical content of the utterance, making it a twofold verification task verifying both the speaker and the phrase.

    In Task 1, each trial includes a test segment, a model identifier indicating three enrollment utterances, and a phrase ID uttered in the utterances. The system processes each trial independently, producing a log-likelihood ratio (LLR) score that combines speaker and phrase verification scores.
    
    Enrollment and test phrases, except for target-wrong trials where some tests were selected from free-text utterances, are drawn from a fixed set of ten phrases, comprising five Persian and five English phrases, respectively. The in-domain training data includes utterances from \textbf{1620 speakers}, some of whom only have Persian phrases. Model enrollment is conducted in a phrase and language-dependent manner, using three utterances for each model. With the same set of target speakers, Task 1 offers a solid basis for analyzing the language factor in text-dependent speaker recognition.
	
	\subsubsection{Trial types}
	
	Given the ground truth, there are four trial types in a TdSV task ~\cite{larcher2014text}. The first is Target-Correct (TC), where the target speaker utters the correct passphrase. The second is Target-Wrong (TW), where the target speaker utters an incorrect passphrase. Similarly, Imposter-Correct (IC) and Imposter-Wrong (IW) refer to cases where the impostor utters the correct or incorrect passphrase. The system should accept TC trials as target trials and reject the other three types as non-target (impostor) trials. Note that the main difference between text-dependent and text-independent speaker verification is the consideration of TW trials as impostor trials, while both TC and TW are considered target trials in text-independent mode. There are no cross-language and cross-gender trials in Task 1 of the challenge.
	
	\subsubsection{Training condition}
	
	The training condition is defined as the amount of data/resources used to build a Speaker Recognition (SR) system. We adopted a fixed training condition where the system should only be trained using a designated set. The fixed training set consists of the following:
	
	\begin{itemize}
		\item VoxCeleb 1\&2
		\item LibriSpeech
		\item Mozilla Common Voice Farsi
		\item DeepMine (Task 1 Train Partition)
	\end{itemize}
	
	The use of other public or private speech data for training is prohibited, while the use of non-speech data for data augmentation purposes is permitted. The in-domain DeepMine training data can be utilized for any purpose, including neural network training, LDA or PLDA model training, and score normalization. Regarding the possibility of using pre-trained models, it should be noted that the use of any public pre-trained model is allowed in this challenge.

    We provide a \textbf{separate development set} for the challenge. Teams can use the development set for parameter tuning and system evaluation before submitting them to the leaderboard to minimize the number of submissions. Any other training, such as fusion training on the development set, is not permitted. Additionally, any usage of Task 2 in-domain data and its development set data for this task is not allowed.
		
	\subsubsection{Enrollment Condition}
	
	Enrollment is conducted using three utterances of a specific phrase for each model. We opted for three utterances for model enrollment, as it is commonly practiced. It's important to note that using enrollment utterances from other models is prohibited, such as for calculating score normalization parameters (i.e., trials are to be processed independently).
	
	\subsubsection{Test Condition}
	
	Each evaluation trial includes a test utterance and a target model. As mentioned earlier, there are four trial types in the evaluation, with only TC considered as the target, and the rest classified as imposture. Among the trial types, IW trials can be easily rejected by a simple verification system. Consequently, there are no IW trials in the trial list.

    Please be aware that there is a limit of 10 submissions per team. Additionally, teams can submit at most two score files per day on the evaluation platform. During the last five days of the challenge deadline, the leaderboard will be hidden.
	
	\subsection{Task 2 - Text-dependent Speaker Verification Using User-defined Passphrases}
	
	Task 2 of the TdSV challenge revisits the problem of speaker verification in a text-dependent manner. This means that with three repetitions of a phrase as enrollment data and a test file, we automatically determine whether the target speaker has uttered the test speech or not. We also need to determine if the uttered phrase in the test speech is the same as the one uttered by the speaker in the enrollment speech. This is the same task addressed by Task 1 of the challenge, with the difference that here we assume the passphrase has been chosen by the user, and we have no training data for it. Additionally, we do not know the specific phrase chosen by the user in terms of content.

    To examine the impact of using additional utterances from the speaker, which are not constrained by any content limitations, for enrollment, in this task, besides the three repetitions of the passphrase chosen by the speaker, there are also other speech files available from the same speaker, in which the uttered phrase is not specified. Teams are allowed to use these utterances in any way they choose to build the speaker model. It is recommended that a comparison be made in the final report between using these utterances and the scenario where these utterances are not used.
	
	Each trial in this task consists of a test speech segment accompanied by a model identifier, indicating three repetitions of the passphrase, as well as several additional free-text enrollment utterances. The system is required to process each trial independently and generate a log-likelihood ratio (LLR) for each of them. The trial types here are the same as in Task 1, meaning TC trials should be accepted, while TW and IC trials should be rejected.

    To simulate user-defined passphrases for this task, four out of the 10 available text-dependent phrases in the dataset are exclusively used in the test set, while the remaining six phrases are retained in the training data. Therefore, the in-domain training data for this task comprises text-independent Persian utterances from 1620 speakers, along with utterances from the aforementioned six phrases. This data can be utilized for various purposes, such as LDA/PLDA, score normalization, training data for neural networks, etc.
	
	\subsubsection{Training Condition}
	
	The training data conditions in this task are similar to Task 1, except that here, specific in-domain data has been included, which is exclusive to this task and should not be utilized for Task 1. Additionally, it is prohibited to use any Task 1 in-domain data and its development set data for this task.
	
	\subsubsection{Enrollment Condition}
	
	As mentioned before, the enrollment data in Task 2 consists of three repetition of the user's passphrase plus one to several variable-length free-text utterances. Teams can use either both phrase dependent and phrase independent utterances for model creation or one of them. Since each enrollment utterance is a complete recording without trimming to a specific duration, the overall duration might not be exactly uniform. Note that using the enrollment utterances from the other models is forbidden, for example, for calculating score normalization parameters.
	
	\subsubsection{Test Condition}
	
	Each evaluation trial comprises a test utterance and a target model. Similar to Task 1, the content of a test utterance should correspond to the user's passphrase; however, here, we assume that the phonetic content of the passphrase is unknown.

    Similarly to Task 1, there is a development set (a smaller evaluation set) with keys that enable teams to monitor performance improvements before submitting scores to the evaluation platform. Additionally, there is a limit of 10 submissions per team, and submitting two score files per day is allowed. During the last five days leading up to the challenge deadline, the leaderboard will be hidden.
	
	\subsection{Performance Measurement for Both Tasks}
	
	The main metric for the challenge is normalized minimum Detection Cost Function (DCF) as defined is SRE08. This detection cost function is defined as a weighted sum of miss and false alarm error probabilities:
	\begin{equation*}
	\label{eq.s_norm}
	C_{Det} = C_{Miss} \times P_{Miss\:|\:Target} \times P_{Target} + C_{FalseAlarm} \times P_{FalseAlarm\:|\:NonTarget} \times (1 - P_{Target})\:,
	\end{equation*}
	where $C_{Miss} = 10$, $C_{FalseAlarm} = 1$ and $P_{Target}=0.01$. Based on the parameters, the normalized DCF ($DCF_{norm}$) will be DCF divide by 0.1 as the best cost that could be obtained without processing the input data. In addition to $minDCF_{norm}^{0.01}$, the Equal Error Rate (EER) will be reported.
	
	\subsection{Data Description}
	
    The primary dataset for the challenge is the DeepMine dataset, collected through crowdsourcing~\cite{lee2015reddots}. Participants in the data collection project installed an Android application and recorded phrases within the application. A comprehensive description of the project and dataset can be found in~\cite{deepmine2019asru, deepmine2018odyssey}. Below are the BibTeX sources for citing the database for easy reference:

	\begin{verbatim}
	@inproceedings{deepmine2018odyssey,
	    title={{DeepMine} Speech Processing Database: Text-Dependent and 
	    Independent Speaker Verification and Speech Recognition in 
	    {Persian and English}.},
	    author={Zeinali, Hossein and Sameti, Hossein and Stafylakis, Themos},
	    year=2018,
	    booktitle={Proc. Odyssey 2018 The Speaker and Language Recognition 
	    Workshop},
	    pages={386--392},
	}
	
	@inproceedings{deepmine2019asru,
	    title={A Multi Purpose and Large Scale Speech Corpus in {Persian and 
	    English} for Speaker and Speech Recognition: the {DeepMine} Database},
	    author={Zeinali, Hossein and Burget, Lukas and Cernocky, Jan},
	    year=2019,
	    booktitle={Proc. ASRU 2019 The 2019 IEEE Automatic Speech Recognition 
	    and Understanding Workshop},
	}
	\end{verbatim}

	The database was recorded in real-life environments in Iran, incorporating various types of noise during collection. The primary language is Persian (Farsi), with most participants also contributing to the English portion. Part 1 of the dataset includes five Persian and five English phrases used in both tasks of the challenge. Table~\ref{tbl.phrases} displays the English phrases and transliterations of the Persian phrases. The phoneme transcription of the phrases are shown in Table~\ref{tbl.transcriptions}, and participants are free to use them as needed. Part 3 of the dataset contains text-independent phrases in Persian and is utilized in the training data, as well as TW trials in the evaluation set.
	
	\begin{table}[tb]
		\renewcommand{\arraystretch}{1.2}
        \label{tbl.phrases}
		\caption{Phrases in Task1 of the challenge.}
		\vspace{-2mm}
		\centerline
		{
			\setlength\tabcolsep{12pt}
			\begin{tabular}{l | l }
				\toprule
				\midrule
				Id & Phrase \\
				\midrule
				01 & sedaye man neshandahandeye hoviyyate man ast. \\
				02 & sedaye har kas monhaser be fard ast. \\
				03 & hoviyyate man ra ba sedaye man tayid kon. \\
				04 & sedaye man ramze obure man ast. \\
				05 & baniadam azaye yekdigarand. \\
				06 & My voice is my password. \\
				07 & OK Google. \\ 
				08 & Artificial intelligence is for real. \\
				09 & Actions speak louder than words. \\
				10 & There is no such thing as a free lunch. \\
				\midrule
				\bottomrule
			\end{tabular}
		}
	\end{table}

    There are 6 vowels in Persian and the following map can be used roughly to map them with English vowels. You can also find another mapping for doing this. Also, you can do a speech analysis to map them.

    \begin{itemize}
        \item Three short vowels:
        \begin{enumerate}
            \item A like "AE" in "rat - R AE T"
            \item E like "EH" in "red - R EH D"
            \item O like "AO" in "short - SH AO R T" but it is pronounced shorter than in English.
        \end{enumerate}
        \item Three long vowels:
        \begin{enumerate}
            \item AA like "AA" in "God - G AA D"
            \item I like "IY" in "seed - S IY D"
            \item U something like "UW" in "too - T UW"
        \end{enumerate}
    \end{itemize}

    \begin{table}[tb]
		\renewcommand{\arraystretch}{1.2}
        \label{tbl.transcriptions}
		\caption{Transcriptions of the challenge phrases. The character "-" is used as a word-separator.}
		\vspace{-2mm}
		\centerline
		{
			\setlength\tabcolsep{12pt}
            \begin{adjustbox}{width=1\textwidth}
			\begin{tabular}{l | l }
				\toprule
				\midrule
				Id & Phrase \\
				\midrule
				01 & S E D AA Y E - M A N - N E SH AA N D A H A N D E Y E - H O V I Y Y A T E - M A N - AH A S T \\
				02 & S E D AA Y E - H A R - K A S - M O N H A S E R - B E - F A R D - AH A S T \\
				03 & H O V I Y Y A T E - M A N - R AA - B AA - S E D AA Y E - M A N - T A AH Y I D - K O N \\
				04 & S E D AA Y E - M A N - R A M Z E - AH O B U R E - M A N - AH A S T \\
				05 & B A N I AH AA D A M - AH A AH Z AA Y E - Y E K D I G A R A N D \\
				06 & M AY - V OY S - IH Z - M AY - P AE S W ER D \\
				07 & OW K EY - G UW G AH L \\ 
				08 & AA R T AH F IH SH AH L - IH N T EH L AH JH AH N S - IH Z - F AO R - R IY L \\
				09 & AE K SH AH N Z - S P IY K - L AW D ER - DH AE N - W ER D Z \\
				10 & DH EH R - IH Z - N OW - S AH CH - TH IH NG - AE Z - EY - F R IY - L AH N CH \\
				\midrule
				\bottomrule
			\end{tabular}
            \end{adjustbox}
		}
	\end{table}

	\subsection{Data Organization}
	
	Data will be provided in separate zip files for each task. Each zip file will include the following components: the in-domain DeepMine training data, enrollment data (for development and evaluation), the model definition file, the trial file (for development and evaluation), and the test utterances. When extracting the files into a directory, the directory structure should be as follows:
	
	\begin{verbatim}
	<base directory>/
	  README.txt
	  docs/
	    dev_model_enrollment.txt
	    dev_trials.txt
	    eval_model_enrollment.txt
	    eval_trials.txt
	    train_labels.txt
	  wav/
	    enrollment/
	      enr_000001.wav
	      enr_000002.wav
	      ...
	    evaluation/
	      evl_000001.wav
	      evl_000002.wav
	      ...
	    train/
	      trn_000001.wav
	      trn_000002.wav
	      ...
	\end{verbatim}
	
	Please note that the zip files for Task 1 and Task 2 should be extracted into separate directories due to partial overlap in their contents.
	
	\subsection{Format of Model Enrollment File}
	
	\subsubsection{Task 1 Enrollment File}
	
	The enrollment file for Task 1 is a space-separated six-column text file named \texttt{model\_enrollment.txt}, located in the \texttt{docs} directory. The file begins with a header line. Each line afterward represents a record. The first column indicates the model ID, the second column shows the phrase ID corresponding to the uttered phrase, the third column indicates the speaker's gender and the remaining three columns display the enrollment file IDs. There is only one space between two records in each line. The format of the enrollment file is as follows:
	
	\begin{verbatim}
	model-id<SPACE>phrase-id<SPACE>gender<SPACE>enroll-file-id1<SPACE>enroll-file-id2
	<SPACE>enroll-file-id3<NEWLINE>
	\end{verbatim}
	where, \texttt{model-id} represents the model identifier, \texttt{phrase-id} denotes the phrase identifier, \texttt{gender} shows the speaker's gender, and \texttt{enroll-file-ids} indicate the identifiers for the enrollment utterances.
	
	For example:
	
	\begin{verbatim}
	model-id phrase-id gender enroll-file-id1 enroll-file-id2 enroll-file-id3
	eval_model_000001 05 m enr_036529 enr_036530 enr_036531
	eval_model_000002 08 f enr_010048 enr_010049 enr_010050
	eval_model_000003 01 f enr_042226 enr_042227 enr_042228
	eval_model_000004 09 f enr_021046 enr_021047 enr_021048
	eval_model_000005 04 f enr_000253 enr_000254 enr_000255
	\end{verbatim}
	
	\subsubsection{Task 2 Enrollment File}
	
	The enrollment file for Task 2 is a space-separated text file named \texttt{model\_enrollment.txt}, located in the \texttt{docs} directory. The file begins with a header line. Each subsequent line contains at least eight columns. The first column indicates the model ID and the second column displays the speaker's gender. The following three columns represent enrollment files containing the speaker's passphrase. The remaining columns contain file identifiers for free text enrollment utterances. The number of columns in each line may vary depending on the number of enrollment files. There is only one space between two columns in each line. The format of the enrollment file is as follows:
	
	\begin{verbatim}
	model-id<SPACE>gender<SPACE>enroll-file-id1<SPACE>enroll-file-id2
	<SPACE>enroll-file-id3<SPACE>enroll-file-id4<SPACE>enroll-file-id5...<NEWLINE>
	\end{verbatim}
	
    In this format, \texttt{model-id} represents the model identifier, \texttt{gender} indicates the speaker's gender, and \texttt{enroll-file-ids} denote the identifiers for the enrollment utterances.
	
	For example:
	
	\begin{verbatim}
	model-id gender enroll-file-ids ...
	eval_model_000001 f enr_014718 enr_014719 enr_014720 enr_014721 enr_014722 
    enr_014723 enr_014687
	eval_model_000002 f enr_039732 enr_039733 enr_039734 enr_039735 enr_015670 
    enr_017664 enr_017668 enr_039736 enr_039737
	eval_model_000003 m enr_052051 enr_052052 enr_052053 enr_006366 enr_006382 
    enr_031688 enr_052054 enr_052055
	eval_model_000004 f enr_045576 enr_045577 enr_045578 enr_045579 enr_041021 
    enr_045580
	eval_model_000005 f enr_013768 enr_013769 enr_013770 enr_013771 enr_013772 
    enr_013773 enr_013774 enr_013775 enr_013776 enr_013777
	\end{verbatim}
	
	\subsection{Format of Trial File}
	
	The trial file for both tasks is a space-separated two-column text file named \texttt{trials.txt}, located in the \texttt{docs} directory. The file begins with a header line. Each subsequent line consists of two columns. The first column indicates the model ID and the second column indicates the evaluation file ID. There is only one space between two columns in each line. The format of the trial file is as follows:
	
	\begin{verbatim}
	model-id<SPACE>evaluation-file-id<NEWLINE>
	\end{verbatim}
	where \texttt{model-id} represents the model identifier, and \texttt{evaluation-file-id} denotes the test utterance identifier.
	
	For example:
	
	\begin{verbatim}
	model-id evaluation-file-id
	eval_model_000001 evl_000122
	eval_model_000001 evl_000186
	eval_model_000001 evl_000548
	eval_model_000001 evl_001205
	eval_model_000001 evl_001416
	eval_model_000001 evl_002123
	eval_model_000001 evl_002129
	eval_model_000001 evl_002230
	\end{verbatim}
	
	\section{In-domain Training Set}
		
	As previously mentioned, the in-domain data for both tasks comprises utterances from 1620 speakers. All training utterances are stored in the \texttt{wav/train} directory. The file \texttt{train\_labels.txt} in the \texttt{docs} directory is a space-separated text file containing information for each utterance. Each line in this file consists of three columns: the first column indicates the \texttt{train-file-id}, the second column denotes the \texttt{speaker-id}, and the last column shows the \texttt{phrase-id}. For training utterances that contain free-text, the \texttt{phrase-id} is labeled as ``FT''. There is a header line at the beginning of the file. The format of the train label file is as follows:
	
	\begin{verbatim}
	train-file-id<SPACE>speaker-id<SPACE>phrase-id<NEWLINE>
	\end{verbatim}
	where \texttt{train-file-id} represents the train utterance identifier, \texttt{speaker-id} indicates the speaker label, and \texttt{phrase-id} is the identifier of the phrase for each utterance.
	
	For example:
	
	\begin{verbatim}
	train-file-id speaker-id phrase-id
	trn_000001 spk_000001 03
	trn_000002 spk_000001 03
	trn_000003 spk_000001 03
	trn_000004 spk_000001 03
	trn_000005 spk_000001 03
	trn_000006 spk_000001 03
	trn_000007 spk_000001 03
	trn_000008 spk_000001 03
	\end{verbatim}

	\section{Evaluation Rules and Requirements}
	
	The overall rules are quite similar to NIST SREs. Firstly, participants must adhere to the data restrictions, where only a fixed training condition is permitted. Participants are also obligated to process the test data according to the following rules and upload the results to the challenge website for evaluation. These rules are:

    \begin{itemize}
        \item Participants agree to make at least one valid submission for one of the tasks.
        \item Participants agree to process each trial independently. That is, each decision for a trial is to be based solely upon the specified test segment and target speaker enrollment data. The use of information about other test segments and/or other target speaker data is not allowed.
        \item Participants agree not to probe the enrollment or test segments via manual/human means, such as listening to the data or producing a manual transcript of the speech.
        \item Participants are allowed to use any automatically derived information for training, development, enrollment, or test segments.
        \item Participants are not allowed to use the development set for any training purposes.
        \item Participants may make multiple submissions for each task. Based on the leaderboard results, participants should select the best-performing systems, write a proper system description based on the results, and send it to challenge organizers via email \url{tdsvc.2024@gmail.com}.
    \end{itemize}
	
	\section{Baseline}
	
	There is a common ECAPA-TDNN baseline for both tasks. 
	
	\section{Prizes}
	
	There will be three cash prizes for each task. Winners will be selected based on the results of the evaluation dataset and other qualitative factors such as the novelty and contribution of the proposed system. Moreover, the results achieved by a single system or fusions of fewer systems rather than a mega fusion are more interesting. In addition to the cash prize, each winner will receive a certificate for their achievement. The cash prizes for each task are as follows:
	\begin{itemize}
		\item Rank 1: \textbf{2000} USD
		\item Rank 2: \textbf{1000} USD
		\item Rank 3: \textbf{500 } USD
	\end{itemize}
	
	\section{Evaluation Protocol}
	
	\subsection{Challenge Website}
	
	The challenge has a GitHub page (\url{https://tdsvc.github.io/}). All information and updates about the challenge will be posted on the website. System descriptions submitted by the participants will also be available on the challenge website.

    \subsection{Registration}

    In summary, to register for this challenge, you first need to have a username in the Codabench system. Then, access the Google form available at the following address, complete the dataset license agreement file, sign it, and send it to us along with completing the form. After reviewing your information, your registration will be completed, and access information to the challenge dataset will be sent to you. You can find more registration details in the challenge on the website.

    \url{https://forms.gle/q2kXMuvomkpG73Gj9}
	
	\subsection{Leaderboard Platform}
	
    As mentioned earlier, there is an online leaderboard for each task, and participants can submit a maximum of two score files per day during the evaluation period. There is also a limit of 10 submissions in total for each task. The leaderboard displays the performance of the systems on the final evaluation set. Please note that 5 days before the challenge deadline, the public leaderboard will be hidden. Teams will be ranked based on the best-performing submitted system.
 
	The challenge leaderboard platforms are available at:
	\begin{itemize}
		\item Task 1:~ \url{https://www.codabench.org/competitions/2760/}
		\item Task 2:~ \url{https://www.codabench.org/competitions/2761/}
	\end{itemize}
	
	\subsection{System Output Format}
	
	The results should be submitted in a single ZIP file containing a single text file named \texttt{answer.txt}. The text file must be located at the root of the ZIP file, and the ZIP file should not contain any folders or other unwanted files.

    The \texttt{answer.txt} file is a one-column text file. Each line of the file indicates an LLR score (a float number) for the corresponding trial. The order of scores must match the order of trials in the trial file, and all trials must be scored. Any inconsistency will result in an error during the evaluation of the system. Please note that while the trial file includes a header line, this header should not be included in the \texttt{answer.txt} file.
    
	For example:
	\begin{verbatim}
	-6.1284
	-97.8528
	-16.8025
	-44.3276
	4.4121
	-61.0123
	-42.9890
	\end{verbatim}

    If the \texttt{answer.txt} file is not found inside the ZIP file or is not correctly formatted as specified above, your submission will likely fail and will NOT be scored. Additionally, improper formatting may lead to the wrong interpretation of your results and therefore to an incorrect score.
	
	\subsection{Data License Agreement}
	
	The evaluation data for this challenge is a subset of the DeepMine dataset. To access this dataset, participants must sign a data license agreement specific to the TdSV Challenge 2024. This license permits participants to use the data for the challenge and subsequent paper publications. Any other use of the data is strictly prohibited. The license agreement file can be found on the challenge website.
	
	\subsection{System Description}
	
	Each participant is required to submit a comprehensive system description, which will be made available online on the challenge website. Additionally, we strongly encourage participants to submit papers to the special session related to the challenge in SLT 2024. These papers will undergo standard review procedures and should adhere to proper formatting and demonstrate sufficient novelty for acceptance.

    The system description must be at least 2 pages long and include the following information about the submitted systems. You can use the INTERSPEECH or SLT paper template for writing the description.
    
    \begin{itemize}
        \item A thorough description of the system components, including front-end and back-end modules, along with their configurations.
        \item A detailed explanation of the data partitions used to train the various models.
        \item Performance metrics of the submitted systems on both the development set and the evaluation set, as reported on the leaderboard website.
    \end{itemize}
    
    A BibTeX source for citing the evaluation plan is provided below for easy reference.
	
	\begin{verbatim}
	@techreport{tdsvc2024plan,
	    title={Text-dependent Speaker Verification (TdSV) Challenge 2024: 
	        Challenge Evaluation Plan.},
	    author={Zeinali, Hossein and Lee, Kong Aik and Alam, Jahangir and 
	    Burget, Luka\v{s}},
	    institution={arXiv preprint arXiv:1xxx.0xxxx},
	    year=2024,
	}
	\end{verbatim}
	
	\section{Planned Evaluation Schedule}
	
	\begin{table}[h]
		\renewcommand{\arraystretch}{1.2}
		\centerline
		{
			\setlength\tabcolsep{20pt}
			\begin{tabular}{l | l }
				Release of Evaluation Plan:                 & Apr 10, 2024 \\
				Release of Train, Dev, and Eval Sets:       & Apr 10, 2024 \\
				Evaluation Platform Open:                   & Apr 20, 2024 \\
				Challenge Deadline:						    & Jun 10, 2024 \\
				System Description Deadline:			    & Jun 17, 2024 \\
				SLT Paper Submission Deadline:			    & Jun 20, 2024 \\
				TdSV Challenge 2024 Special Session at SLT:	& Dec 02 – 05, 2024 \\
			\end{tabular}
		}
	\end{table}
	
	\bibliographystyle{IEEEbib}
	\bibliography{main}
	
\end{document}